# Anomalous scaling and spin-charge separation in coupled chains

Peter Kopietz, Volker Meden, and Kurt Schönhammer
*Institut für Theoretische Physik der Universität Göttingen,*
*Bunsenstr.9, D-37073 Göttingen, Germany*
(19 August 1994)

We use a bosonization approach to show that the three dimensional Coulomb interaction in coupled metallic chains leads to a Luttinger liquid for vanishing inter-chain hopping $t_\perp$, and to a Fermi liquid for any finite $t_\perp$. However, for small $t_\perp \neq 0$ the Greens-function satisfies a homogeneity relation with a non-trivial exponent $\gamma_{cb}$ in a large intermediate regime. Our results offer a simple explanation for the large values of $\gamma_{cb}$ inferred from recent photoemission data from quasi one-dimensional conductors and might have some relevance for the understanding of the unusual properties of the high-temperature superconductors.

In a recent Letter [1] Chakravarty and Anderson pointed out that their interlayer-tunneling theory of high-temperature superconductivity [2] is based on the fundamental assumption that the single-particle Greens-function $G(\mathbf{k}, \omega)$ in a sufficiently large regime of wave-vectors $\mathbf{k}$ and frequencies $\omega$ satisfies the homogeneity relation $G(s\mathbf{k}, s\omega) = s^{\gamma-1}G(\mathbf{k}, \omega)$, with an exponent $\gamma > 0$. It is well known that the Greens-function of a Luttinger liquid (LL) satisfies such a scaling law. In a Fermi liquid (FL), however, the anomalous exponent $\gamma$ vanishes. There exists now general agreement that for regular interactions the LL exists only in dimensions $d = 1$. To stabilize the LL in higher dimensions, it is necessary to invoke singular interactions [3], the existence of which in realistic experimental systems remains controversial [4]. A generally accepted microscopic justification for the assumption of anomalous scaling made by Chakravarty and Anderson does not exist. In this Letter we offer a possible justification.

We proceed in three steps: We first consider a highly anisotropic metal consisting of an array of chains embedded in a three dimensional lattice. Assuming that the electrons interact via Coulomb forces, we show that for vanishing inter-chain hopping $t_\perp$ the system is a LL, and calculate the anomalous exponent $\gamma_{cb}$. At parameters relevant in recent photoemission experiments [5], we find that $\gamma_{cb}$ is of the order of unity. As a second step, we consider the same system with $t_\perp \neq 0$, assuming that $t_\perp$ is small compared with the Fermi energy $E_F$. We show that, in spite of the fact that for any finite $t_\perp$ the system is a FL, the Greens-function for small enough $|t_\perp/E_F|$ exhibits anomalous scaling in a large intermediate regime, *with the same exponent $\gamma_{cb}$ as for $t_\perp = 0$*. Hence, for sufficiently small inter-chain hopping the experiments indeed measure the properties of the LL-fixed point. In *this sense $t_\perp$ is an irrelevant* perturbation. We argue that our result is to a large extent model independent, and that there exists at least a whole class of Fermi liquids with small quasi-particle residue $Z \ll 1$ which exhibits the characteristic LL-features in a large intermediate regime of momenta and frequencies. To strengthen this hypothesis, we finally consider an exactly solvable one-dimensional FL with exponentially small $Z$, and show that the spectral function of this model is practically indistinguishable from that of a LL.

We start with an anisotropic $3d$ metal consisting of a quadratic array of $1d$ chains where $a_\perp$ is the distance between the chains. As the fundamental force between electrons in a metal is determined by the Coulomb potential $e^2/|\mathbf{r} - \mathbf{r}'|$, we do not replace the bare interaction by hand by an effective short-range interaction, but solve the screening problem explicitly. For a *single* chain the one-dimensional Fourier transform $U(q_\parallel)$ requires a regularization at short distances, which will be discussed below. For small $q_\parallel$ one obtains a logarithmic singularity in $U(q_\parallel)$ such that the anomalous dimension diverges and the system is *not* a LL [6]. The *Coulomb coupling between the chains* removes this singularity and the three dimensional system without hopping between the chains is a LL [7].

For the individual chains in the high density limit we can ignore scattering processes with large momentum transfer and linearize the energy dispersion around the two Fermi points. Then without hopping between the chains the electron number in each chain is conserved separately and we have a problem of coupled Tomonaga-Luttinger models (TLM) which can e.g. be solved with the help of a canonical transformation [8]. As we later address also the problem including transverse hopping we use here the results of the bosonization approach for dimensions $d \geq 1$ [9–11]. The expression for the one-particle Greens-function of a given chain involves the RPA screened interaction

$$U^{RPA}(q_\parallel, z) = \left\langle \frac{U(q_\parallel, \mathbf{q}_\perp)}{1 + \Pi_0(q_\parallel, z)U(q_\parallel, \mathbf{q}_\perp)} \right\rangle \quad , \quad (1)$$

where the parenthesis $< \ldots >$ denotes the average over the two-dimensional first Brillouin zone (BZ), i.e.

$< \ldots > = (\frac{a_\perp}{2\pi})^2 \int_{-\pi/a_\perp}^{\pi/a_\perp} d\mathbf{q}_\perp \ldots$. Here $\Pi_0(q_\parallel, z) = (2\hbar v_F q_\parallel^2/\pi)/[(\hbar v_F q_\parallel)^2 - z^2]$ is the $1d$ non-interacting density response-function, where $v_F$ is the Fermi velocity. $U(q_\parallel, \mathbf{q}_\perp)$ is the mixed $2d$-lattice and $1d$-continuous Fourier transform of the Coulomb potential,

$$U(q_\parallel, \mathbf{q}_\perp) = e^2 \int dx \sum_{\mathbf{r}_\perp} \frac{e^{-i[q_\parallel x + \mathbf{q}_\perp \cdot \mathbf{r}_\perp]}}{\sqrt{x^2 + \mathbf{r}_\perp^2}} \quad . \quad (2)$$

The $\mathbf{r}_\perp = 0$ term we regularize by assuming a Gaussian form of the one electron wave function perpendicular to the chains with a decay length $a_0$ [12]. As the effective interaction involves the $1d$ density response function, the one-particle Greens-function has a form very similar to the single chain model [13,14]. If $\alpha = +(-)$ denotes right (left) moving electrons, the part of the Greens-function describing photoemission can be written as $G_\alpha^<(x,t) = [G_\alpha^<(x,t)]^{(0)} \exp[Q_\alpha^<(x,t)]$, where $[G_\alpha^<(x,t)]^{(0)}$ is the non-interacting Greens-function, and $Q_\alpha^<(x,t) = S_\alpha^<(0,0) - S_\alpha^<(x,t)$, with

$$S_\alpha^<(x,t) = -\frac{1}{2}\frac{2\pi}{L}\sum_{q_\parallel > 0}\frac{1}{q_\parallel}\left\{e^{-i\alpha q_\parallel x}\left[<e^{i\tilde{v}_F(\mathbf{q})q_\parallel t}>\right.\right.$$
$$\left.\left.-e^{iv_F q_\parallel t}\right] + \cos(q_\parallel x)<2s^2(\mathbf{q})e^{i\tilde{v}_F(\mathbf{q})q_\parallel t}>\right\} \quad , \quad (3)$$

where $L$ is the length of the chains, $\tilde{v}_F(\mathbf{q}) = v_F[1 + 2U(q_\parallel, \mathbf{q}_\perp)/(\pi\hbar v_F)]^{\frac{1}{2}}$, and $s^2(\mathbf{q}) = [\tilde{v}_F(\mathbf{q}) - v_F]^2/[4\tilde{v}_F(\mathbf{q})v_F]$. The only difference to the single-chain result is the average $< \ldots >$ over the BZ. The relevant dimensionless coupling constant follows from Eqs. (2) and (3) as $g = e^2/(\pi\hbar v_F)$. The anomalous dimension $\gamma_{cb}$ is given by $\gamma_{cb} = \frac{1}{2}\lim_{q_\parallel \to 0} < 2s^2(\mathbf{q}) >$, and is diverging for a single chain with Coulomb forces [6]. In order to discuss the result for coupled chains we have to evaluate $U(q_\parallel, \mathbf{q}_\perp)$. For arbitrary $\mathbf{q}_\perp$ this has to be done numerically where we use an Ewald summation technique [15]. For $|\mathbf{q}_\perp|a_\perp \ll 1$, we may approximate $U(q_\parallel, \mathbf{q}_\perp) \approx 4\pi e^2/(a_\perp \mathbf{q})^2$, the usual $3d$ Fourier transform of the Coulomb potential. As mentioned before, the $\mathbf{r}_\perp = 0$ contribution in Eq. (2) has a term $-c\ln(q_\parallel a_0)$, diverging for $q_\parallel \to 0$. This divergence is *exactly cancelled* by a term $c\ln(q_\parallel a_\perp)$ from the $\mathbf{r}_\perp \neq 0$ part of the sum, leaving a contribution $c\ln(a_\perp/a_0)$. For every finite value of $|\mathbf{q}_\perp|$ therefore $s^2(\mathbf{q})$ tends to a finite value for $q_\parallel \to 0$ and the system is a LL.

The momentum integrated spectral function $\rho(\omega)$, which apart from a one-electron dipole matrix element determines angular integrated photoemission, is algebraically suppressed near the chemical potential $\rho(\omega) \propto |\omega|^{\gamma_{cb}}$. Recent photoemission studies of quasi-one dimensional conductors suggest values for the anomalous dimension in the range $1.0 \pm 0.2$ [5]. This would be hard to reconcile with a model involving short range interactions, as e.g. the anomalous dimension in the one dimensional Hubbard model never exceeds $1/8$. Our treatment using real Coulomb forces shows that this quite easily leads to anomalous dimensions in the experimental range, when the value of the dimensionless coupling $g$ is estimated for these systems [16]. In Fig. 1 we show results for $\gamma_{cb}$ as a function of $g$ for two values of $a_\perp/a_0$ using the numerical evaluation of the sum in Eq. (2). As a comparison we also show the result when $U(q_\parallel, \mathbf{q}_\perp)$ is replaced by $4\pi e^2/(a_\perp \mathbf{q})^2$, which provides the correct description for $g|\ln(a_\perp/a_0)| \ll 1$. In this limit the averages over the BZ can be evaluated analytically for $t = 0$ [17]. This shows explicitly that the BZ averaging in Eq. (3) leads to an ultraviolet cutoff for the $q_\parallel$-summation which in this limit is given by the Thomas-Fermi wave-vector $\kappa = a_\perp^{-1}[8\pi g]^{\frac{1}{2}} = a_\perp^{-1}[8e^2/(\hbar v_F)]^{\frac{1}{2}}$ and can be determined numerically in the more realistic parameter regimes.

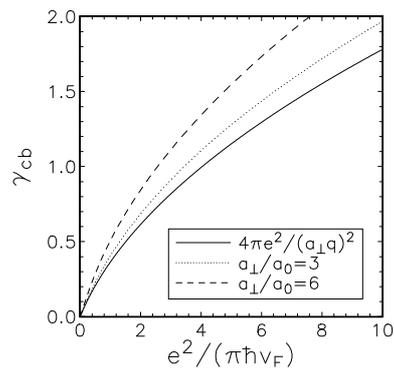

FIG. 1. The anomalous dimension $\gamma_{cb}$ as a function of the coupling constant for two values of $a_\perp/a_0$ (dashed and dotted lines) and $U(q_\parallel, \mathbf{q}_\perp) = 4\pi e^2/(a_\perp \mathbf{q})^2$ (solid line).

In order to evaluate the momentum dependent spectral function $\rho^<(q_\parallel, \omega)$ we have used the numerical techniques developed in Ref. [14]. It is not obvious from Eq. (3) if a charge peak occurs as the charge-velocity $\tilde{v}_F(\mathbf{q})$ depends on $\mathbf{q}_\perp$ and is subject to the BZ averaging. Because the spin velocity $v_F$ is independent of $\mathbf{q}_\perp$ this leads for $\gamma_{cb} < \frac{1}{2}$ to a sharp threshold singularity at $\omega = -v_F|q_\parallel|$, as in the $1d$ TLM [14,18]. For $\gamma_{cb} > \frac{1}{2}$ the singularity is washed out, but the threshold survives. To clearly demonstrate spin-charge separation, we therefore present in Fig. 2 spectra for $\gamma_{cb} = 0.407$, corresponding to $g = 1$.

Experimentally the inter-chain hopping $t_\perp$ can never be completely turned off. *In the renormalization group sense*, $t_\perp$ is a relevant perturbation and leads to a finite quasi-particle residue $Z > 0$ [19]. We now use the bosonization approach in $d > 1$ [9–11] to show that for small $\theta = |t_\perp/E_F|$ the behavior of the Greens-function at experimentally measurable length scales is completely dominated by the LL-exponent. Thus, *from the experimental point of view, $t_\perp$ is an irrelevant perturbation*. A crucial step in the bosonization approach is the subdivision of the Fermi surface into a set of disjoint patches $K_\alpha$.

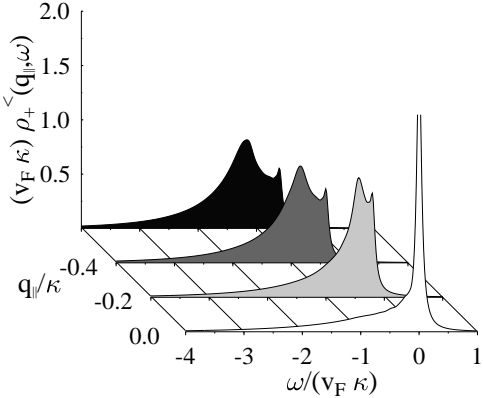

FIG. 2. The Lorentzian broadened spectral function $\rho_+^<$ for the coupled chains as a function of $\omega$ for different $q_\parallel$. The parameters are $g = 1$, $a_\perp/a_0 = 3$ and the broadening $\chi/(v_F \kappa) = 0.05$. $\omega$ is measured relative to $\mu$ and $q_\parallel$ relativ to $k_F$.

For $t_\perp = 0$ these patches can be identified with the two sheets of the Fermi surface, $K_\pm = \{k_x = \pm k_F; |k_y|, |k_z| < \pi/a_\perp\}$, where we have assumed that the chains lie in the $x$-direction. It is not necessary to divide $K_\pm$ into smaller patches, because the velocities $\mathbf{v}_\alpha = \nabla_\mathbf{k}\epsilon_\mathbf{k}|_{\mathbf{k}\in K_\alpha} = \pm\hbar v_F \hat{\mathbf{k}}_x$ are constant on $K_\pm$. Here $\hat{\mathbf{k}}_x$ is a unit vector in the $x$-direction. Thus, Eq. (3) can be obtained as a special case of the general bosonization formula for the Greens-function in $d > 1$, which for arbitrary patch topology has been written down in Ref. [11]. In contast to the $t_\perp = 0$ calculation, this approach is approximate in character. In the quasi one-dimensional materials discussed in Ref. [16] the intra-chain hopping $t_x$ is an order of magnitude larger than the inter-chain hopping $t_y = t_\perp$, which is again a factor of ten larger than $t_z$. We therefore assume transverse hopping only in the $y$-direction. In this case the sheets $K_\pm$ acquire a small modulation along the $k_y$-direction, with amplitude proportional to $\theta k_F$. To describe this modulation in terms of patches, let us assume that the band structure is such that the distortion of patches $K_\pm$ can be approximated by wedges with hight $\theta k_F$ and peak on the $k_x$-axis. The Fermi surface can then be described by four patches $K_\alpha$, $\alpha = 1, \ldots, 4$, on which the velocities $\mathbf{v}_\alpha = [\pm\hat{\mathbf{k}}_x\cos\theta \pm \hat{\mathbf{k}}_y\sin\theta]\hbar v_F$ are constant. Using the general bosonization formula for the Greens-function in an arbitrary patch-geometry [11], we find after some lengthy but straight-forward algebra [17] that for $0 < \theta \ll 1$

$$R_\alpha \equiv \lim_{L\to\infty} S_\alpha^<(0,0) = -\gamma_{cb}\left[\ln\theta^{-1} + c_1\right] \quad , \qquad (4)$$

and for $1 \ll \tilde{\kappa}|r_\parallel| \ll \theta^{-1}$

$$S_\alpha^<(r_\parallel, 0) = -\gamma_{cb}\left[\ln(\theta\tilde{\kappa}|r_\parallel|)^{-1} + c_2\right] \quad , \qquad (5)$$

where $c_1$ and $c_2$ are numerical constants of the order of unity, and $r_\parallel^2 = r_x^2 + \theta^2 r_y^2$ [20]. Here $\tilde{\kappa}$ is an ultraviolet cutoff, which for small $g|\ln(a_\perp/a_0)|$ can be identified with the Thomas-Fermi wave-vector $\kappa$, and otherwise is in the range between $a_\perp^{-1}$ and $a_0^{-1}$. The important point is now that *the prefactor of the logarithmic terms in Eqs. (4) and (5) is given by the anomalous dimension $\gamma_{cb}$ of the LL that would exist for $\theta = 0$*. Because $R^\alpha$ is finite, the system is a FL [11], with quasi-particle residue $Z = e^{R_\alpha} \propto \theta^{\gamma_{cb}}$. Thus, $Z$ vanishes as a non-universal power-law as $\theta \to 0$, with the exponent given by the anomalous dimension $\gamma_{cb}$ of the corresponding LL. The transition LL $\to$ FL is due to the symmetry-breaking of the shape of the Fermi surface [9], which induces a *coupling of the phase space* associated with the $q_x$- and $q_y$-integrations. Combining Eqs. (4) and (5), we see that $Q_\alpha^<(r_\parallel, 0) = -\gamma_{cb}[\ln(\tilde{\kappa}|r_\parallel|) + c_1 - c_2]$. This is precisely the characteristic LL-behavior, so that for $1 \ll \tilde{\kappa}|r_\parallel| \ll \theta^{-1}$ the equal time Greens-function is of the LL-form and satisfies the scaling law $G_\alpha(\mathbf{r}/s, 0) = s^{3+\gamma_{cb}}G_\alpha(\mathbf{r}, 0)$, with an exponent $\gamma_{cb}$ that is determined by the corresponding LL at the same value of the coupling constant.

The finite-$t$ behavior of $S_\alpha^<(r_\parallel, t)$ cannot be obtained analytically. However, we expect that anomalous scaling holds also at finite $t$, and that the appearance of characteristic LL-features in our system is a model-independent property of a whole class of Fermi liquids with small quasi-particle residue. A detailed numerical study of this point for our anisotropic Coulomb-model with $t_\perp \neq 0$ will be presented in Ref. [17].

To give further evidence in favour of the generality of our observation, we now briefly discuss a very different FL with $Z \ll 1$, and show that it has similar properties as our anisotropic Coulomb-model. Suppose we replace in the TLM the usual dimensionless couplings $g_2$ and $g_4$ by $F_q = g(q/\kappa)^\eta e^{-q/\kappa}$, with $0 < \eta \ll 1$, $g > 0$, and $\kappa \ll k_F$. The result for $G_\alpha^<(x, t)$ is of the same form as in Eq. (3), except that the BZ averages should be omitted, and $\tilde{v}_F(\mathbf{q})$ should be replaced by $\tilde{v}_F(q) = v_F[1 + 2F_q]^{\frac{1}{2}}$. While for $\eta = 0$ the $q \to 0$ limit of $F_q$ is finite and leads to a logarithmic divergence of $S_\alpha^<(0,0)$ as $L \to \infty$ and LL behavior, this divergence is suppressed for $\eta > 0$ and one obtains a FL with quasi-particle residue $Z = e^{R_\alpha}$. Obviously, this FL is obtained from the TLM by deforming the potential energy, whereas the distortion of the shape of the Fermi surface in our three dimensional model discussed above is a kinetic energy effect. To examine if the Greens-function satisfies anomalous scaling in some parameter regime, we calculate

$$G_\alpha^<(x/s, t/s) = s[G_\alpha^<(x, t)]^{(0)}Ze^{-S_\alpha^<(x/s, t/s)} \quad . \qquad (6)$$

We find for the quasi-particle residue $Z = e^{-\gamma(g)/\tilde{\eta}(g)}$, and for $S_\alpha^<(x, 0)$ in the large distance limit $|\kappa x| \gg 1$

$$S_\alpha^<(x, 0) = -\frac{\gamma(g)}{\tilde{\eta}(g)|\kappa x|^{\tilde{\eta}(g)}} \quad , \qquad (7)$$

where to leading order in $\eta$ the exponent $\gamma(g)$ can be identified with the anomalous dimension of the corresponding TLM with coupling constants $g_2 = g_4 = g$, i.e. $\gamma(g) = \frac{1}{4}[(1+2g)^{\frac{1}{2}} - 1]^2/(1+2g)^{\frac{1}{2}}$. The exponent $\tilde{\eta}(g)$ is a featureless monotonically decreasing function of $g$, varying between $\tilde{\eta}(0) = 2\eta$ and $\tilde{\eta}(\infty) = \eta/2$. Note that for any $g > 0$ and sufficiently small $\eta$ the quasi-particle residue becomes exponentially small. It is now trivial to prove the existence of anomalous scaling in an exponentially large regime. Writing $|\kappa x|^{-\tilde{\eta}} = e^{-\tilde{\eta} \ln |\kappa x|}$ and assuming $\tilde{\eta}|\ln|\kappa x|| \ll 1$, Eq. (7) implies to leading order $S_\alpha^<(x/s, 0) \approx S_\alpha^<(x, 0) - \gamma(g) \ln s$. Using then Eq. (6), we conclude that $G_\alpha^<(x/s, 0) \approx s^{1+\gamma(g)} G_\alpha^<(x, 0)$ in the exponentially large regime $1 \ll |\kappa x| \ll e^{1/\tilde{\eta}(g)}$. The restriction due to the lower limit holds also in a LL, but in the FL anomalous scaling breaks down at $x_{max}(g) = \kappa^{-1} e^{1/\tilde{\eta}(g)}$. Obviously the anomalous scaling regime exists only for $\eta \ll 1$. Note also that at strong coupling $x_{max}(g)$ is much larger than at weak coupling, in agreement with the intuitive picture that strong correlations should lead to more pronounced anomalous properties.

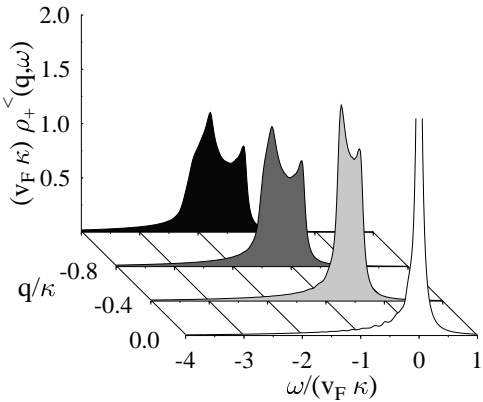

FIG. 3. The Lorentzian broadened spectral function $\rho_+^<$ for the $1d$ FL as a function of $\omega$ for different $q$. The parameters are: $\eta = 0.1$, $g = 2$ and $\chi/(v_F \kappa) = 0.05$. The small oscillations are a finite-size effect.

At finite $t$ we have used again the technique of Ref. [14] to calculate the spectral function. In Fig. 3, we present the spectral function of the FL for $\eta = 0.1$, which is to the accuracy of the drawing indistinguishable from the spectral function of a TLM with $g_2 = g_4 = g$ [21]. Because we know that the spectral function of the TLM satisfies anomalous scaling, we conclude for $\eta > 0$ the FL must have this property as well. From Fig. 3 it is obvious that our FL exhibits spin-charge separation in the scaling regime.

Our result provides a very simple microscopic justification for the assumption of anomalous scaling made by Chakravarty and Anderson [1] to explain the unusual properties of the high-temperature superconductors. Assuming that due to strong correlations and anisotropy these systems are Fermi liquids with small quasi-particle residue, we predict that they should exhibit all the characteristic LL-features in the experimentally accessible range of wave-vectors and frequencies. Thus, although dimensionality and transverse hopping are relevant in the renormalization group sense, they can be considered as irrelevant as far as the experiments are concerned.

One of us (K.S.) would like to thank D. Baeriswyl and G. M. Eliashberg for stimulating discussions.